\documentclass[%
reprint,
superscriptaddress,
 amsmath,amssymb,
 aps,
prl,
longbibliography,
]{revtex4-2}
\usepackage{textcomp}
\usepackage{graphicx}% Include figure files
\usepackage{dcolumn}% Align table columns on decimal point
\usepackage{bm}% bold math
\usepackage{color}
\usepackage{lineno}
\begin{document}

\preprint{APS/123-QED}

\title{
Magnetic Permeability Time-varying Metamaterials at Microwave Frequencies
}
% Force line breaks with \\

\author{Toshiyuki Kodama}
\email[Email address:]{tkodama@tohoku.ac.jp}
\affiliation{Institute for Excellence in Higher Education, Tohoku University, Sendai, 980-8576, Japan}

\author{Nobuaki Kikuchi}
\affiliation{Department of Mathematical Science and Electrical-Electronic-Computer
Engineering, Graduate School of Engineering, Akita University, Akita 010-8502, Japan}

\author{Takahiro Chiba}
\affiliation{Frontier Research Institute for Interdisciplinary Sciences, Tohoku University, Sendai 980-8578, Japan}
\affiliation{Department of Applied Physics, Graduate School of Engineering, Tohoku University, Sendai 980-8579, Japan}

\author{Seigo Ohno}
\affiliation{Department of Physics, Graduate School of Science, Tohoku University, Sendai 980-8578, Japan}

\author{Satoshi Okamoto}
\affiliation{Institute of Multidisciplinary Research for Advanced Materials, Tohoku University, Sendai 980-8577, Japan}
\affiliation{Center for Science and Innovation in Spintronics, Tohoku University, Sendai 980-8577, Japan}

\author{Satoshi Tomita}
\email[Email address:]{tomita@tohoku.ac.jp}
\affiliation{Institute for Excellence in Higher Education, Tohoku University, Sendai, 980-8576, Japan}
\affiliation{Department of Physics, Graduate School of Science, Tohoku University, Sendai 980-8578, Japan}

\date{\today}% It is always \today, today,

\begin{abstract}
We demonstrate magnetic permeability ($\mu$) time-varying metamaterials 
at GHz frequencies using ferromagnetic permalloy (Ni$_{\rm 80}$Fe$_{\rm 20}$; Py). 
We observe frequency up and down conversion 
of 4 GHz microwaves through the metamaterials, 
which is caused by the temporal modulation 
of $\mu$ in the Py layer.
Moreover, 
the efficiency of the up-conversion to a higher frequency 
is much larger 
than that of the down conversion to a lower frequency.
These experimental results are reproduced well 
via numerical calculation, 
verifying that 
the significant up-conversion efficiency 
is traced back to nonlinear magnetization dynamics 
in the metamaterials.
The present study opens a door to 
microwave sources toward the 6th-generation mobile communication system, 
four-dimensional metamaterials with spatio-temporal modulation, 
and nonlinear spintronics.
\end{abstract}

%\keywords{Suggested keywords}%Use showkeys class option if the keyword
\maketitle

\textit{Introduction} --
Noether's theorem \cite{Noether} says that 
a system with a continuous symmetry 
has a conserved quantity. 
In a medium with the space-translational symmetry,
the conjugated quantity, wavevector of light, is conserved.
When the space-translational symmetry is broken 
at a spatial boundary between different indices of refraction, 
for example, at the interface of a prism,
the wavevector is not conserved, 
resulting in refraction of light \cite{Hecht}. 
Contrastingly,
frequency (energy) of light is 
a conjugated quantity 
in a medium with the time-translational symmetry. 
In a time-varying medium 
with broken time-translation symmetry,
therefore, 
frequencies of light have to change
while leaving the wavevector unchanged \cite{Miyamaru2021-ph}; 
this is the principle of frequency conversion 
using an electro-optic (EO) modulator
toward optical frequency-comb in modern optics \cite{Udem2002, Ishizawa2023}.
This principle is applicable in man-made structured materials, 
metamaterials, 
exhibiting exotic optical properties unavailable in natural materials.

The most typical metamaterials are 
metamaterials with a negative index of refraction \cite{Shelby2001-rn} 
and for invisible cloaks \cite{Schurig2006-kk}. 
These are referred to as space-varying metamaterials 
because the wavevectors of light are changed 
at the spatial boundary.
Recently, 
a new paradigm for light generation and steering 
has emerged \cite{Galiffi2022-kc} -- 
time-varying metamaterials, 
in which refractive indices are temporally modulated. 
Time-varying metamaterials
would bring about promising applications 
to frequency converters \cite{Zhang2022-ei, Taravati2021-wm, Zhou2020-of},
nonreciprocal devices \cite{Guo2019-vq}, 
and anti-reflection temporal coatings \cite{Pacheco-Pena2020-yq, Liberal2023-ld}
as well as exotic phenomena, 
for example, 
time refraction \cite{Zhou2020-of} and reflection \cite{Apffel2022-go, Moussa2023-pv}, 
analogue of a continuous time crystal \cite{Liu2023-nw}, 
and temporal aiming \cite{Pacheco-Pena2020-jn}.

The temporal modulation of refractive indices 
has been achieved experimentally 
using time modulated electric permittivity ($\varepsilon$) \cite{Miyamaru2021-ph,Zhou2020-of}. 
However, 
time modulation of magnetic permeability ($\mu$) is expected 
because $\mu$ is the counterpart of $\varepsilon$ 
in the refractive index $n = \sqrt{\varepsilon} \sqrt{\mu}$.
The space-time modulation of both $\varepsilon$ and $\mu$
is essential for investigating 
Fresnel drag for light in optically moving media \cite{Huidobro2019-eg}, 
spatio-temporal four-dimensional control of wave-matter interactions \cite{Engheta2023-ta}, 
quantum electrodynamics phenomena \cite{Kort-Kamp2021-pa, Silveirinha2023-ip, Pendry2024-pb}, 
and Doppler cloaks \cite{Liu2020-rk}.
In any cases, 
the key is the realization of $\mu$ time-varying metamaterials. 
Nevertheless, 
experimental demonstration of time modulation of $\mu$ is lacking.

\begin{figure*}[tb!]
\includegraphics[width=18truecm, clip]{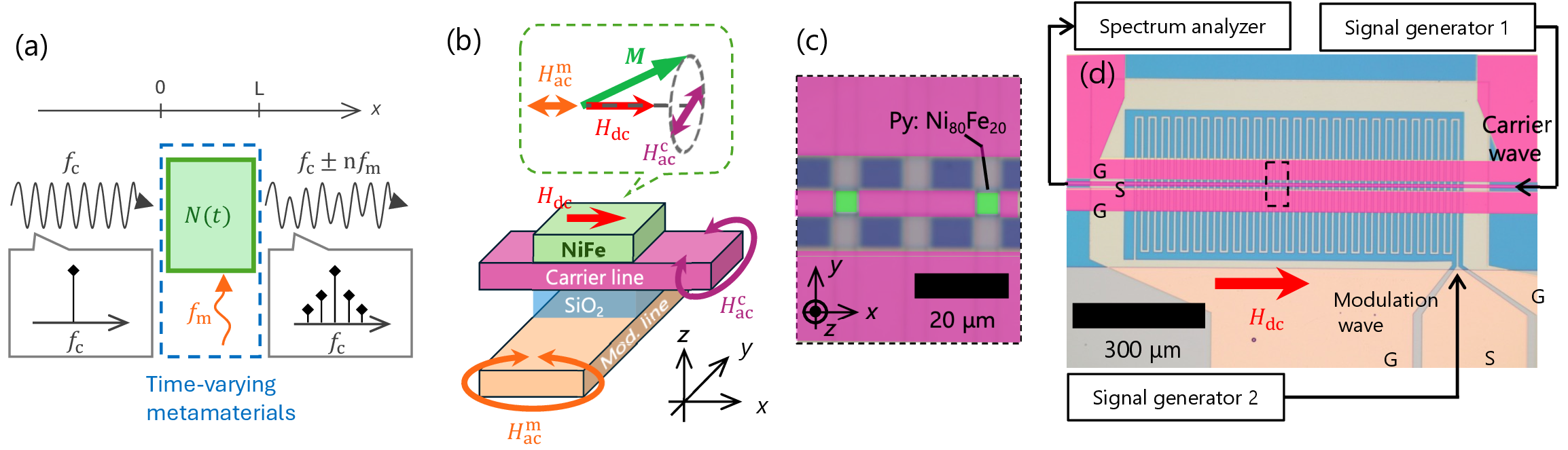}
\caption{
(a) Schematic illustration of frequency conversion 
by a time-varying metamaterial with a refractive index $N (t)$
and a modulation frequency $f_{\rm m}$. 
A frequency of a carrier wave to be modulated is $f_{\rm c}$.
(b) Bottom: schematic illustration of the metamaterial structures.
Top: configurations of magnetization ($M$) in precession and magnetic fields ($H$) 
for excitation and modulation of FMR.
(c) Enlarged optical microscope image of the metamaterial.
(d) Overview photo of the metamaterial 
together with illustration of measurement setup. 
The enclosed area by the black dashed line corresponds to (c).
}
\label{sample}
\end{figure*}

In an electrical circuit analogue,
$\mu$ modulation is regarded as an inductance variation.
As inductance is relevant to turn numbers, length, and cross section of coils, 
time-varying $\mu$ intuitively seems to be a tough challenge. 
However, 
$\mu$ can be varied in magnetic materials.
In particular, 
ferromagnetic materials 
show frequency dispersion of $\mu$ 
in the vicinity of ferromagnetic resonance (FMR) frequencies 
in the microwave region.
Furthermore, 
a large magnetization 
originated from coupled microscopic electron spins 
in the ferromagnets 
gives rise to a large variation in $\mu$.
In this way, 
$\mu$ time-varying metamaterials at GHz frequencies
are realized
when the FMR conditions are temporally modulated.
Although several possibilities using spintronic techniques 
are proposed \cite{Kodama2023PRAppl, Kodama2024PRB}, 
an important issue to be addressed is 
how the effective magnetic fields for FMR are dynamically changed 
at high frequencies. 

In this Letter,
we prepare magnetic metamaterials 
including a ferromagnetic-metal (permalloy (Py), Ni$_{\rm 80}$Fe$_{\rm 20}$) layer, 
together with microwave carrier and modulation lines. 
While the microwaves in the carrier line 
excites FMR in the Py layer,
a time-periodic Oersted magnetic field 
generated by another microwave in the modulation line 
alters the effective magnetic field for FMR 
temporally at GHz frequencies.
We observe frequency conversion 
of the carrier microwaves 
and verify that 
the conversion is caused by the temporal modulation 
of $\mu$ in the Py layer on FMR.
A magnetic counterpart of the EO modulator 
is realized at microwave frequencies.  
The most striking feature 
of the $\mu$ time-varying metamaterial 
is that the up-conversion efficiency 
is much larger 
than the down-conversion efficiency.
This would be advantageous 
to study Floquet engineering of light-matter interaction \cite{Oka2019, Yin2022} 
and realize microwave/millimeter-wave sources 
toward the 6th-generation mobile communication system.  

The experimental results are reproduced well 
via numerical calculation, 
indicating that 
the significant up-conversion efficiency 
is traced back to nonlinear magnetization dynamics 
in the $\mu$ time-varying metamaterials.
Although nonlinear magnetization dynamics 
and microwave response through nonlinear susceptibility 
have attracted much attention in spintronics \cite{Sharma2017-gp, Wu2024-co}, 
the physical mechanism is still unclear.
The present study is thus 
an important step for further investigation in the nonlinear magnonics \cite{Zheng2023}.
Furthermore, 
as the present temporally $\mu$ modulated metamaterials 
are prepared on microwave circuits, 
it can be easily integrated with $\varepsilon$-modulated metamaterials.
Therefore,
the present study paves a way for space-time modulation 
of both $\varepsilon$ and $\mu$ in metamaterials, 
opening a door for experimental demonstration of 
relativistic and quantum effects 
in spatio-temporal four-dimensional metamaterials 
\cite{Huidobro2019-eg, Engheta2023-ta, Kort-Kamp2021-pa, Silveirinha2023-ip, Pendry2024-pb}.

\textit{Principle of frequency conversion by permeability modulation} --
Figure \ref{sample}(a)
shows 
a schematic illustration of 
frequency conversion 
by a time-varying metamaterial.
Suppose a one-dimensional propagation 
of a carrier wave having a frequency of $f_{\rm c}$ 
into the $x$ direction. 
The carrier wave is injected to 
a medium 
with a time-dependent complex refractive index $N(t) = n(t) - j \kappa(t)$, 
which is modulated in time by a modulation wave 
having a frequency of $f_{\rm m}$.
The length of the medium $L$ is much smaller than the carrier wavelength.
The plane wave's magnetic field at time $t$ and position $L$ in the medium, 
$H(t, x=L)$, 
is expressed as
\begin{subequations}
\label{emw}
\begin{align}
H(t, x = L) & =H_{0} {\rm exp}{[ 2\pi j (f_{\rm c} t - \frac{N(t) f_{\rm c}}{c} L)]}, \\
& = H_{0} {\rm exp} (-2 \pi \frac{\kappa(t) f_{\rm c}}{c} L)
{\rm exp}{[ 2\pi j (f_{\rm c} t - \frac{n(t) f_{\rm c} }{c} L)]} ,
\end{align}
\end{subequations}
where
$H_{0}$ is an amplitude of the carrier wave and $c$ is the speed of light in vacuum.
Equation (\ref{emw}b) indicates that
the first exponential term containing $\kappa(t)$ 
corresponds to amplitude modulation.
Contrastingly, 
the second exponential term including $n(t)$
represents phase modulation.
When $n(t)$ is varied periodically as sin$(\omega t)$,
this term is described using the Vessel's function.
The carrier wave frequency is thus converted as 
$f_{\rm c} \pm {\rm n} f_{\rm m}$ with ${\rm n} = 1, 2, \cdots$.
Note that 
even if 
the medium originally
shows a linear response, 
for example, FMR in a ferromagnetic material,
an energy 
injected by the modulation wave with $f_{\rm m}$
causes 
nonlinear processes, 
resulting in 
frequency conversions of the carrier waves.
In this way,
a linear medium with the applied modulation power, 
surrounded by blue dashed lines in Fig. \ref{sample}(a),
is regarded as 
the time-varying metamaterial.

Time modulated $N(t)$ can be realized
through $\varepsilon(t)$ and $\mu(t)$.
In this study, 
$\mu(t)$ is achieved using
ferromagnetic metamaterials.
The bottom part of Fig. \ref{sample}(b) 
illustrates the structure of the $\mu(t)$ metamaterial's unit cell.
The metamaterial consists of four layers:
a ferromagnetic Py top layer (green) 
with magnetization ($M$) 
under dc magnetic field $H_{\rm dc}$, 
a gold (Au) carrier line (red)
for FMR excitation by ac magnetic field $H_{\rm ac}^{\rm c}$, 
a silicon dioxide (SiO$_2$) spacer (blue),
and an Au modulation line (orange)
for generating ac magnetic field $H_{\rm ac}^{\rm m}$ 
to periodically change the FMR condition and vary the permeability of Py.

The upper panel of Fig. \ref{sample}(b)  
presents 
the configuration of magnetization $M$ 
and magnetic fields $H$
in the Py layer.
The carrier microwave current 
with a frequency of $f_{\rm c}$
injected into the carrier microstrip line
generates
an ac Oersted magnetic field $H_{\rm ac}^{\rm c}(t)$ 
by the Amp\`{e}re's law.
Under $H_{\rm dc}$, 
the ac Oersted field $H_{\rm ac}^{\rm c}$, 
which is along the $y$-axis and perpendicular to $H_{\rm dc}$,
drives precession of in-plane $M$ in the Py layer (FMR). 
The FMR condition 
is described by
the Kittel's equation for thin magnetic films as
\begin{equation}
\label{kittel}
2 \pi f_{\rm FMR} 
= 
\gamma \sqrt{\mu_{0} H_{\rm ext} (\mu_{0} H_{\rm ext} + \mu_{0} M_{\rm eff})}.
\end{equation}
The FMR frequency $f_{\rm FMR}$ 
is determined by
the gyromagnetic ratio $\gamma$,
the effective magnetic fields along precession axis $H_{\rm ext}$, 
the effective magnetization $M_{\rm eff}$,
and magnetic permeability of vacuum $\mu_0$.

The modulation microwave current 
with a frequency of $f_{\rm m}$
generates
another ac Oersted magnetic field $H_{\rm ac}^{\rm m}(t)$
around the modulation microstrip lines.
As in Fig. \ref{sample}(b), 
$H_{\rm ac}^{\rm m}$ oscillating in $f_{\rm m}$ 
is parallel to $H_{\rm dc}$ 
so that the effective magnetic fields for FMR ($H_{\rm ext}$) 
is expressed as
$H_{\rm ext}(t) = H_{\rm dc} + H_{\rm ac}^{\rm m}(t) $.
In this way,
time modulation of $H_{\rm ext}$ 
with the frequency of $f_{\rm m}$
results in 
dynamic change in $f_{\rm FMR}$.

The relative permeability $\mu_{\rm r}$ 
of the Py layer on FMR
shows frequency dispersion
at the vicinity of $f_{\rm FMR}$.
The real $\mu_{\rm r}'$ and 
imaginary $\mu_{\rm r}''$ parts of $\mu_{\rm r}$ 
at the carrier wave frequency $f_{\rm c}$
are written as \cite{Kodama2024PRB}
\begin{subequations}
\label{mur}
\begin{align}
\mu_{\rm r}' 
&=
1 +  \frac{\gamma \mu_{0} M_{\rm eff}}{2\pi} \frac{f_{\rm FMR} (f_{\rm FMR}^{2} - f_{\rm c}^{2}) + f_{\rm FMR} f_{\rm c}^{2} \alpha^{2}}
{[f_{\rm FMR}^{2} - f_{\rm c}^{2} (1 + \alpha^{2})]^{2} + 4 f_{\rm FMR}^{2} f_{\rm c}^{2} \alpha^{2}} , \\
\mu_{\rm r}''
&= 
\frac{\gamma \mu_{0} M_{\rm eff}}{2\pi} 
\frac{\alpha f [f_{\rm FMR}^{2} - f_{\rm c}^{2} (1 + \alpha^{2})]}
{[f_{\rm FMR}^{2} - f_{\rm c}^{2} (1 + \alpha^{2})]^{2} + 4 f_{\rm FMR}^{2} f_{\rm c}^{2} \alpha^{2}},
\end{align}
\end{subequations}
where 
$\alpha$ is the Gilbert damping parameter.
Equations (\ref{mur}a) and (\ref{mur}b) indicate that, 
when $f_{\rm FMR}$ is dynamically changed 
in time with the frequency of $f_{\rm m}$, 
the time modulation of $\mu_{\rm r}'$ and $\mu_{\rm r}''$ 
is plausible.

\textit{Experimental procedures} --
Figures \ref{sample}(c) and \ref{sample}(d)
present
optical microscopy images 
of the metamaterial 
consisting of arrays of the unit cell 
illustrated in Fig. \ref{sample}(b).
Fig. \ref{sample}(c) corresponds to 
the area enclosed by the black dashed line in Fig. \ref{sample}(d).
The metamaterial is prepared 
by thin films deposition on an undoped silicon substrate 
using Ar-ion magnetron sputtering at room temperature 
followed by microfabrication 
using photolithography and lift-off techniques.
Figure \ref{sample}(c) shows that  
the vertical Au modulation lines (orange) 
have width of 5 {\textmu}m and spacing of 10 {\textmu}m. 
The modulation line
is prepared 
as a meandering track repeated 25 times,
of which the total length is 16 mm,
as shown in Fig. \ref{sample}(d).
The Au thickness 
of the modulation line 
is 200 nm.
An SiO$_2$ layer (blue) of 200 nm thickness 
is deposited
followed by the deposition of another Au layer 
and patterning of the horizontal carrier line (red).
The carrier line is patterned as a coplanar waveguide (CPW) 
consisting of a 5 {\textmu}m-width signal line (S) 
sandwiched between two 50 {\textmu}m-width ground lines (G). 
The gap between S and G lines 
is 5 {\textmu}m. 
Eventually,
a 5-nm-thick tantalum (Ta) layer
followed by a 250-nm-thick Py layer
are deposited 
at the 5 $\times$ 5 {\textmu}m intersection area (green)
of the modulation and carrier lines. 
Because
the current flows in the opposite direction 
between  the adjacent modulation lines, 
the Ta/Py layers are formed 
at every other intersection
as shown in the Fig. \ref{sample}(c).
The total number of the unit cell is 25 
so that the total length of the metamaterial is 725 {\textmu}m.

Figure \ref{sample}(d)
illustrates the measurement setup.
A dc magnetic field 
$\mu_{0} H_{\rm dc}$ in a range of $+$70 mT to $-$70 mT 
is applied along the $x$-axis using an electromagnet.
One side of the CPW
is connected to a signal generator 
via a 10 dBm attenuator 
for the carrier microwave current injection.
The signal generator power is set to 18 dBm so that 
the actual power for FMR excitation is 8 dBm, 
corresponding to $P_{\rm c} =$ 6.31 mW 
if a resister of 50 $\Omega$ is connected instead of the CPW.
The carrier wave with $P_{\rm c} =$ 6.31 mW 
generates the amplitude of $H_{\rm ac}^{\rm c}$ 
to be 1.15 mT at the Py patch.
(See Fig. S1 in Supplemental Materials \cite{SM}).
The injected carrier wave frequency $f_{\rm c}$ is fixed at 4.0 GHz.
The wavelength of the carrier waves in vacuum (75 mm)
is much larger than
the length of 
the unit cell (5 {\textmu}m) and
the metamaterial (725 {\textmu}m).
The metamaterial is thus excited uniformly by the carrier waves.

The spectra of transmitted carrier waves are measured 
using a spectrum analyzer via a 10 dBm attenuator.
The signals measured by the spectrum analyzer 
are averaged by 10$^4$ times. 
The settings of the spectrum analyzer are as follows: 
the center frequency of 4.0 GHz,
the measurement span of 4.0 GHz,
the number of data points of 20001,
and the resolution bandwidth of 3 MHz.
Another signal generator for temporal modulation
is connected to the meandering modulation line.
Microwaves with $f_{\rm m}$ and 20 dBm, 
corresponding to $P_{\rm m} =$ 100 mW, 
are injected into the modulation line.
The evaluated $H_{\rm ac}^{\rm m}$ 
of the 100 mW modulation waves
is 4.6 mT at the Py patch.
(See Fig. S1 in Supplemental Materials \cite{SM}). 
The modulation frequency 
$f_{\rm m}$ is varied from 0.1 GHz to 1.8 GHz.

\begin{figure}[tb!]
\includegraphics[width=8.5truecm,clip]{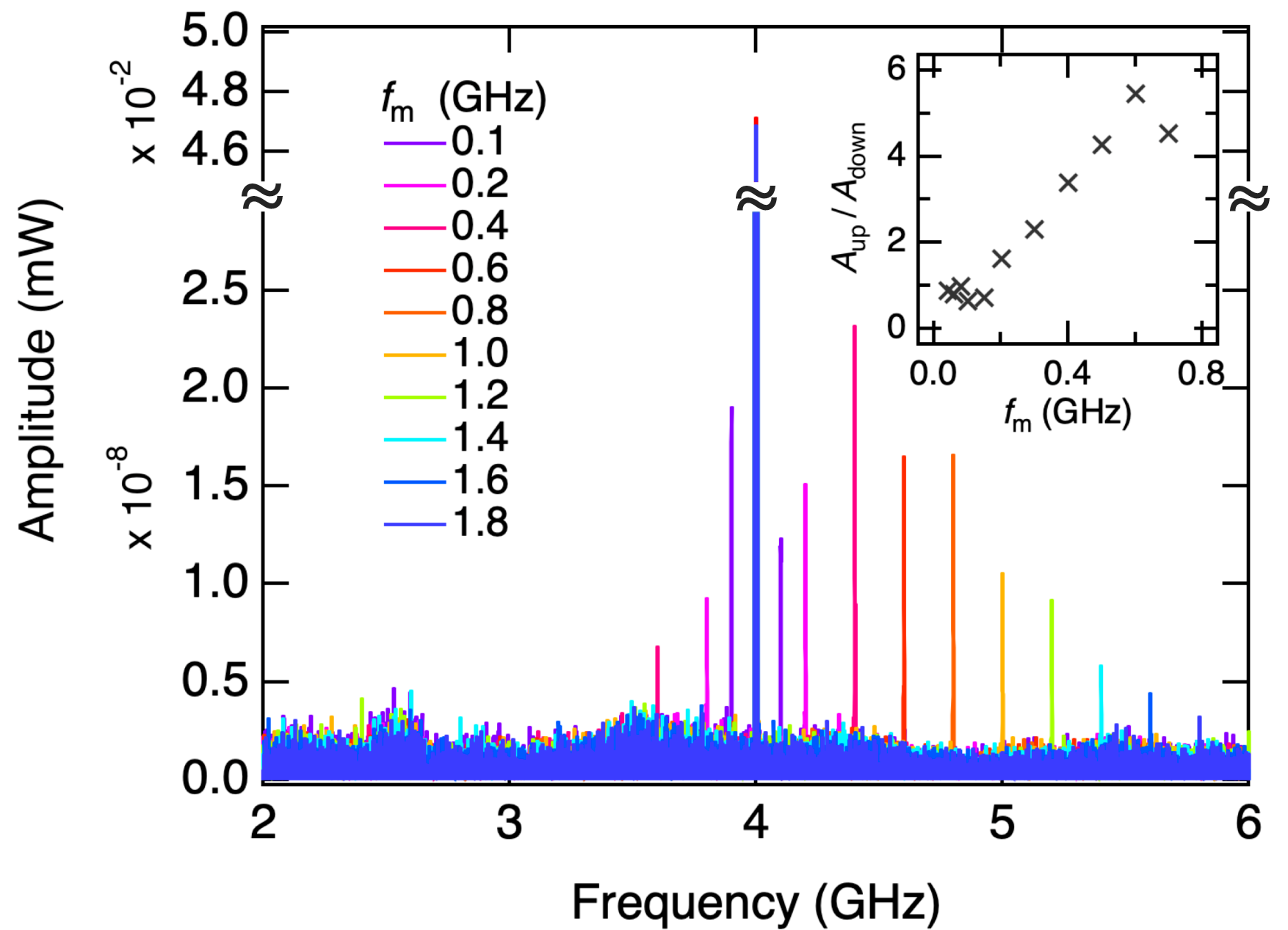}
\caption{
Output microwaves spectra 
of $f_{\rm c} =$ 4.0 GHz 
under $\mu_{0} H_{\rm dc}$ = 24.4 mT
at various modulation frequency $f_{\rm m}$ 
in the range from 0.1 GHz to 1.8 GHz.
Inset: peak amplitude ratio 
$A_{\rm up} / A_{\rm down}$
is plotted as a function of $f_{\rm m}$.
}
\label{conversion}
\end{figure}

\textit{Experimental results} --
Figure \ref{conversion}
shows spectra of the transmitted microwaves 
of $f_{\rm c} =$ 4.0 GHz
at various $f_{\rm m}$ 
under $\mu_{0}H_{\rm dc}$ = 24.4 mT.
The vertical axis 
represents the microwave power.
The most powerful signal at 4.0 GHz 
is an unmodulated carrier wave.
When $f_{\rm m}$ = 0.1 GHz (purple), 
two signals 
are observed besides the 4.0 GHz carrier microwave 
with a spacing of $f_{\rm m}$ = 0.1 GHz: 
signals at a lower frequency of 3.9 GHz 
and at a higher frequency of 4.1 GHz, 
corresponding to down- and up-converted waves 
of the $f_{\rm m}$ = 4.0 GHz carrier wave, 
respectively.
The frequencies of the converted waves 
are expressed as 4.0 $\pm f_{\rm m}$ GHz.
At $f_{\rm m}$ up to 0.4 GHz (pink), 
both the down- and up-converted waves are observed 
in Fig. \ref{conversion}.
The down-converted wave
at $f_{\rm m} =$ 0.6 GHz (orange) 
is detectable 
but the intensity is comparable to the noise.
At above $f_{\rm m}$ = 0.8 GHz (red), 
the down-converted wave cannot be detected. 

Note that 
Fig. \ref{conversion} 
presents only the first-order converted signals.
However, 
the second-order converted signals
expressed as $f_{\rm c} \pm 2 f_{\rm m}$ 
are observed in high sensitivity measurements 
with the measurement span of 1 MHz,
number of data points of 2001,
and resolution bandwidth of 9.1 kHz.
Indeed in Fig. S2 in Supplemental Materials \cite{SM},
high sensitivity measurements 
with $f_{\rm c} =$ 4.0 GHz and $f_{\rm m}$ = 0.1 GHz 
demonstrate the $f_{\rm c} \pm 2 f_{\rm m}$ signals at 3.8 and 4.2 GHz
although the signal intensities are very small.

Figure \ref{conversion} presents that, 
at $f_{\rm m}$ = 0.1 GHz,
the peak amplitude of the down conversion signal $A_{\rm down}$
is larger than that of the up conversion signal $A_{\rm up}$.
However, 
$A_{\rm up}$ becomes more intensive  
than $A_{\rm down}$
as $f_{\rm m}$ increases from 0.2 to 1.8 GHz. 
This is remarkable   
because the $A_{\rm down}$ and $A_{\rm up}$ 
should be the same according to Eq. (\ref{emw}).
In the inset of Fig. \ref{conversion}, 
the amplitude ratio $A_{\rm up} / A_{\rm down}$
is plotted as a function of $f_{\rm m}$ from 0.04 to 0.7 GHz.
Whereas 
$A_{\rm up}$
is slightly smaller than $A_{\rm down}$ 
at $f_{\rm m} \leq$ 0.18 GHz, 
$A_{\rm up}$
becomes larger than $A_{\rm down}$ 
at $f_{\rm m} \geq $ 0.2 GHz.
Furthermore, 
as $f_{\rm m}$ increases,
$A_{\rm up}/A_{\rm down}$
increases monotonically;
eventually at $f_{\rm m} = $ 0.6 GHz,
$A_{\rm up}$ is 5.5 times larger than $A_{\rm down}$.

\begin{figure}[tb!]
\includegraphics[width=8.5truecm,clip]{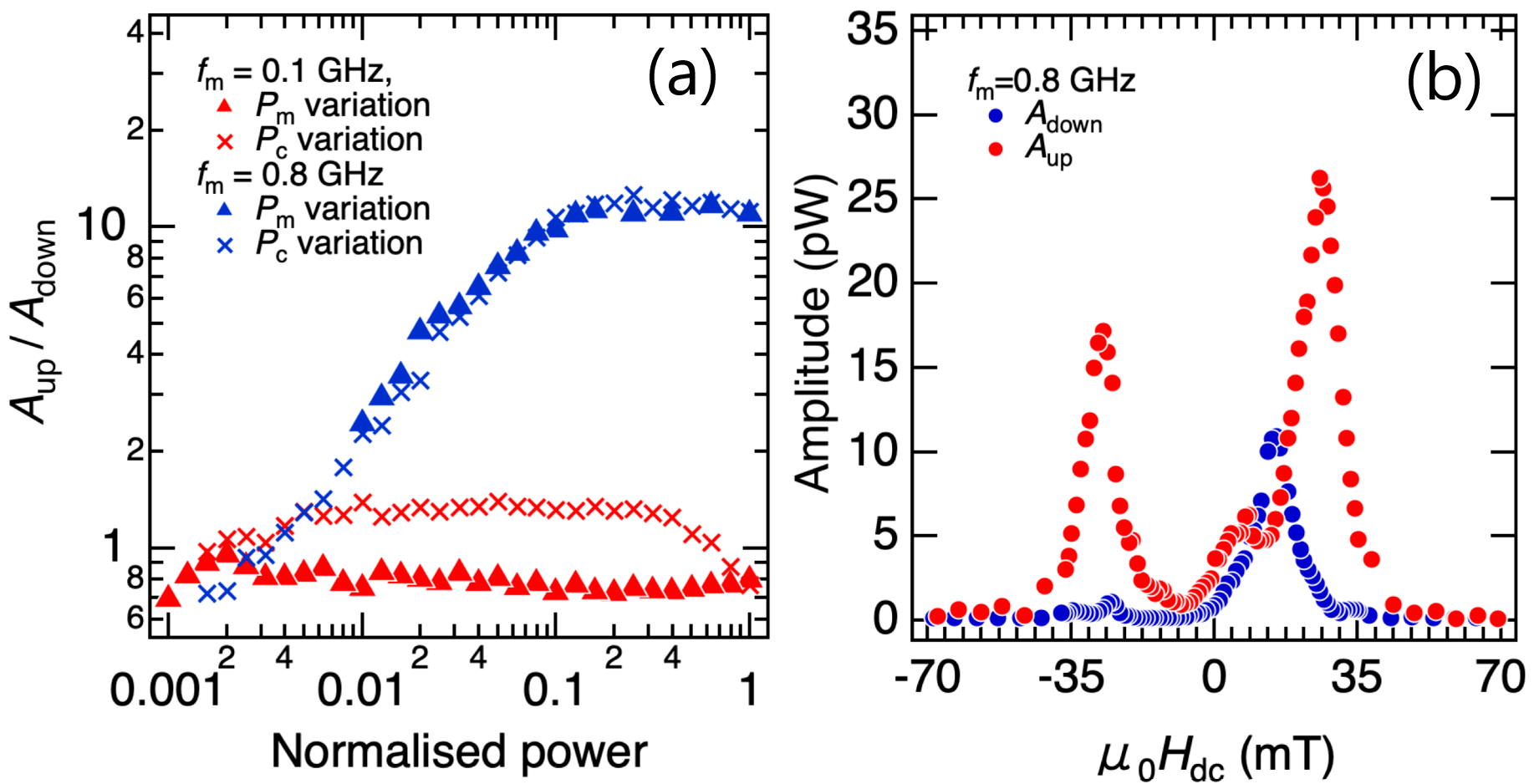}
\caption{
(a) Peak amplitude ratios of up-conversion to 
down-conversion signals $A_{\rm up}/A_{\rm down}$ 
at $\mu_{0} H_{\rm dc} =$ 24.4 mT
are plotted as a function of 
normalized modulation power $P_{\rm m} / (100 \: \rm{mW})$ (triangles) 
and carrier power $P_{\rm c} / (6.31 \: \rm{mW})$ (crosses).
Red and blue correspond to $f_{\rm m} = $ 0.1 GHz and $f_{\rm m} = $ 0.8 GHz, 
respectively.
(b) $A_{\rm down}$ at 3.2 GHz (blue) and $A_{\rm up}$ at 4.8 GHz (red)
with $f_{\rm m}$ = 0.8 GHz 
are plotted as a function of $\mu_{0} H_{\rm dc}$.
}
\label{result}
\end{figure}

The dependence of $A_{\rm up}/A_{\rm down}$ on 
modulation power $P_{\rm m}$ and carrier power $P_{\rm c}$ 
is studied in high sensitivity measurements
as $P_{\rm m}$ or $P_{\rm c}$ decreases.
Triangles in Fig. \ref{result}(a) present 
$A_{\rm up} / A_{\rm down}$ 
at $\mu_{0} H_{\rm dc} =$ 24.4 mT
as a function of normalized modulation power 
$P_{\rm m} / (100 \: \rm{mW})$.
When $f_{\rm m} = $ 0.1 GHz (red triangles),
$A_{\rm up}/A_{\rm down}$
is approximately 1 and is independent of $P_{\rm m}$.
In contrast,
when $f_{\rm m} = $ 0.8 GHz (blue triangles),
$A_{\rm up}/A_{\rm down}$ remains constant at 10 
as $P_{\rm m} /(100\: \rm{mW})$ decreases to 0.2,
and $A_{\rm up}/A_{\rm down}$ decreases monotonically
as $P_{\rm m} /(100\: \rm{mW})$ further decreases to 0.002.
Crosses in Fig. \ref{result}(a) 
highlight 
dependence of $A_{\rm up}/A_{\rm down}$ 
on the normalized carrier power $P_{\rm c} / (6.31 \: \rm{mW})$.
At $f_{\rm m} = $ 0.1 GHz (red crosses) and
$f_{\rm m} = $ 0.8 GHz (blue crosses), 
$A_{\rm up}/A_{\rm down}$ depends on $P_{\rm c}$ 
in a similar way with $P_{\rm m}$. 
Given that 
the vertical and horizontal axes of Fig. \ref{result}(a) are logarithmic scales, 
$A_{\rm up}/A_{\rm down}$ at $f_{\rm m} = $ 0.8 GHz (blue)
shows nonlinear response to $P_{\rm m}$ and $P_{\rm c}$.

In Fig. \ref{result}(b), 
when $f_{\rm c}$ = 4.0 GHz and $f_{\rm m}$ = 0.8 GHz,
$A_{\rm up}$ at 4.8 GHz (red circles)
and $A_{\rm down}$ at 3.2 GHz (blue circles) 
are plotted as a function of 
$\mu_{0} H_{\rm dc}$ from $+$70 to $-$70 mT. 
As $\mu_{0} H_{\rm dc}$ decreases from $+$70 mT,
both $A_{\rm up}$ and $A_{\rm down}$ increase.
While
$A_{\rm up}$ reaches the maximum at $\mu_{0} H_{\rm dc} =$ 25.7 mT,
$A_{\rm down}$ reaches the maximum at $\mu_{0} H_{\rm dc} =$ 15.2 mT.
These magnetic fields for the maximum amplitude 
are consistent with the Kittel's FMR condition.
(See Fig. S3(c) in Supplemental Materials \cite{SM} in more detail.)
With a further decrease in $\mu_{0 }H_{\rm dc}$ to 0 mT,
the amplitude $A_{\rm up}$ and $A_{\rm down}$ decreases to zero.
In the $\mu_{0} H_{\rm dc} < 0$ region,
$A_{\rm down}$ at 3.2 GHz 
reaches the maximum at $\mu_{0} H_{\rm dc} =$ $-$25.2 mT,
while $A_{\rm up}$ at 4.8 GHz 
shows the maximum at $\mu_{0} H_{\rm dc} =$ $-$27.2 mT.
Furthermore, 
the amplitude difference 
between $A_{\rm up}$ at $\mu_{0} H_{\rm dc} > 0$ 
and $A_{\rm up}$ at $\mu_{0} H_{\rm dc} < 0$ 
is traced back to the magnetization hysteresis 
on vortex states \cite{Wurft2019-zu}.
(See Fig. S4 in Supplemental Materials \cite{SM}).

\textit{Theoretical consideration} --
To shed light on the physical origin 
of the asymmetric frequency conversion efficiency, 
we theoretically consider 
a forced oscillation model
based on the Maxwell's equation for electromagnetic waves
combined with the Landau-Lifshitz-Gilbert equation for magnetization precession.
In the model, 
a microwave magnetic field 
is weakly coupled to 
the magnetization dynamics of the magnetic material.
See the calculation details in Supplemental Material \cite {SM}
and references \cite{Chiba2024-in, Mita2024} therein.
As an initial condition, 
an FMR excitation field of 4.0 GHz
with an amplitude of 4.6 mT  
and 
an ac modulation field of $f_{\rm m}$ 
with an amplitude of 1.15 mT 
are considered.
The dc magnetic field $\mu_0 H_{\rm dc}$ is 25 mT.
The magnetization precession 
is driven at 4.0 GHz 
and modulated at $f_{\rm m}$.
$f_{\rm m}$ is varied from 0.2 GHz to 2.0 GHz.
The microwave magnetic field $\tilde{h}_{\rm ac}^{\rm c}$
is calculated in the time domain. 
The time-axis waveforms of $\tilde{h}_{\rm ac}^{\rm c}$
are Fourier transformed to the spectra.  

\begin{figure}[tb!]
\includegraphics[width=8.5truecm,clip]{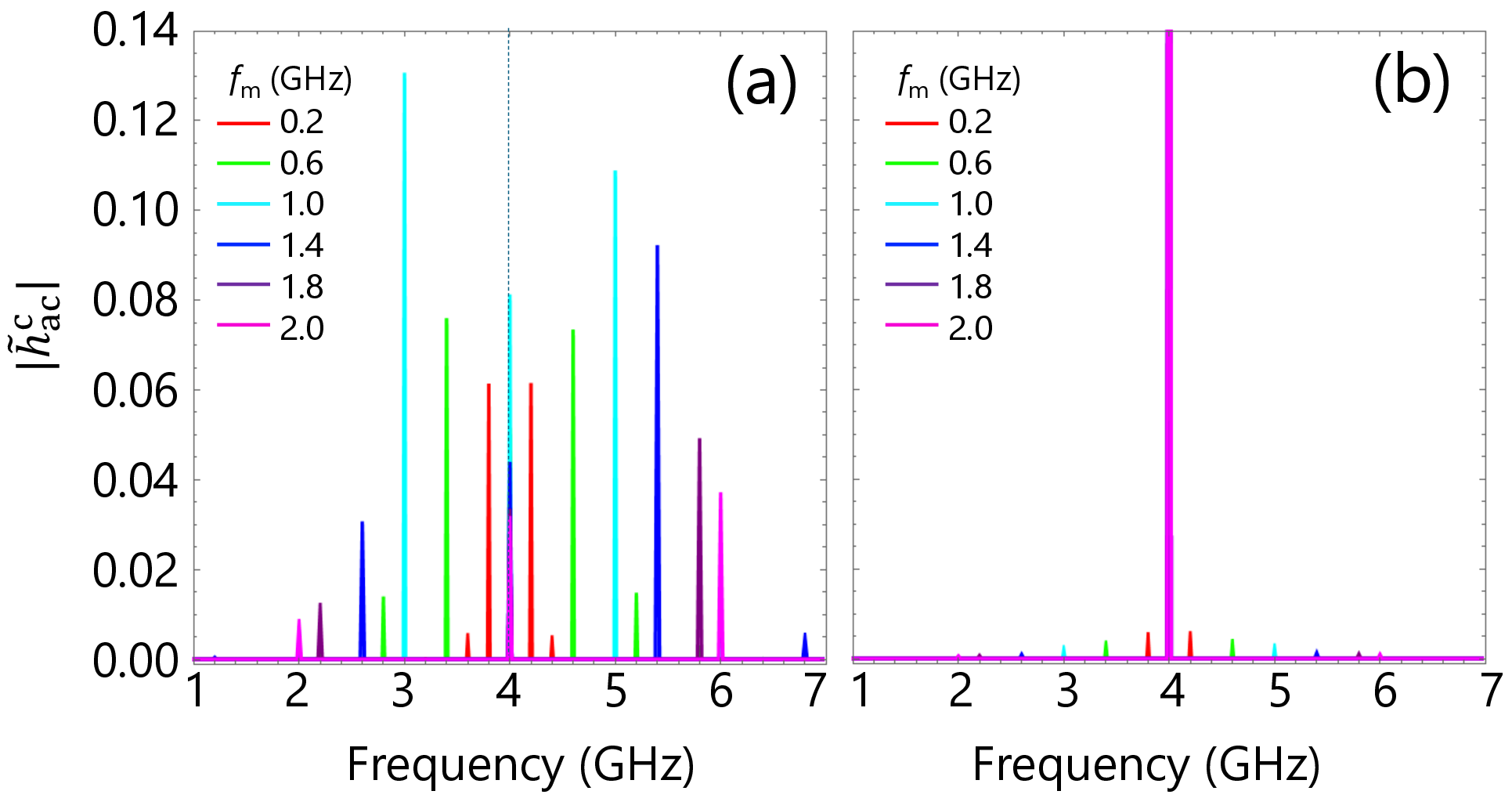}
\caption{
Calculation results of microwave magnetic field $\tilde{h}_{\rm ac}^{\rm c}$
(a) with and (b) without the demagnetization field 
arising from the shape anisotropy.
While $f_{\rm c}$=4.0 GHz is fixed 
as indicated by the vertical dashed line, 
$f_{\rm m}$ is varied from 0.2 GHz to 2.0 GHz.
}
\label{calc}
\end{figure}

Figure \ref{calc}(a)
presents the calculated spectra of $|\tilde{h}_{\rm ac}^{\rm c}|$
affected by a demagnetization field of the thin film shape of Py.
Up- and down-converted signals 
at  $f_{\rm c} \pm f_{\rm m}$
is reproduced well
in the numerical calculation.
At $f_{\rm m}$ = 0.2 GHz (red) and 0.6 GHz (green), 
amplitudes of $|\tilde{h}_{\rm ac}^{\rm c}|$ of the up- and down-converted signals 
are similar.
Note that the second-order converted signals are observed 
at 4.0 $\pm 2 f_{\rm m}$ GHz in Fig. \ref{calc}(a).
At $f_{\rm m}$ = 1.0 GHz (light blue), 
$|\tilde{h}_{\rm ac}^{\rm c}|$ of the up-conversion signal 
is slightly smaller than that of the down-conversion signal.
However, 
when $f_{\rm m}$ is further increased to 1.4 GHz (blue), 1.8 GHz (purple), and 2.0 GHz (magenta),
$|\tilde{h}_{\rm ac}^{\rm c}|$ of the up-conversion signal
becomes significantly larger 
than that of the down-conversion signal. 
The relative $|\tilde{h}_{\rm ac}^{\rm c}|$
reproduces qualitatively 
experimental results observed in the inset of Fig. \ref{conversion}.
The quantitative discrepancy
between the calculation and experimental results
is probably originated from the magnetic dipole interactions 
of a micron-scale sample, which are not taken into account in our macrospin mode.

Figure \ref{calc}(b) 
shows calculation results of a model without demagnetization field.
When the demagnetization field is zero,
corresponding to the Py sphere,
$|\tilde{h}_{\rm ac}^{\rm c}|$ of converted signals 
decrease drastically  in Fig. \ref{calc}(b).
Moreover, 
at $f_{\rm m}$ = 2.0~GHz,
$A_{\rm up}/A_{\rm down}$
of 4.45 in Fig. \ref{calc}(a) 
decrease to 1.63 in Fig. \ref{calc}(b).
Therefore,
the demagnetization field 
due to the magnetic shape anisotropy 
causes nonlinear magnetization dynamics \cite{Zheng2023}, 
giving rise to asymmetrical frequency conversions.

\textit{Discussion} --
In Fig. \ref{result}(b), 
$A_{\rm up}$ and $A_{\rm down}$ 
as a function of $\mu_{0} H_{\rm dc}$ 
highlights that 
no conversion is observed  
at $\mu_{0} H_{\rm dc} = 0$ 
and
$\pm$ 70 mT.
Moreover, 
up- and down-conversion efficiency
shows maximum value 
when $\mu_{0} H_{\rm dc}$ is 
in FMR condition of each converted wave.
These results clearly indicate that 
the frequency conversion is
caused by the time modulation in $\mu$ of Py on FMR.
In this way,
we have demonstrated
$\mu$ time-varying metamaterials
at microwave frequencies.

The most striking feature 
in the microwave frequency conversion 
by the $\mu$ time-varying metamaterials 
is that the up-conversion efficiency 
is increased in a nonlinear fashion and  
much larger than the down-conversion efficiency
particularly at larger $f_{\rm m}$ 
as shown experimentally in Fig. \ref{conversion} 
and numerically in Fig. \ref{calc}.
Experimentally at $f_{\rm m} =$ 0.6 GHz, 
$A_{\rm up}$ is 5.5 times larger than $A_{\rm down}$.
Equations (\ref{emw}a) and (\ref{emw}b)
under moderate modulation fields
do not predict such asymmetric conversion.
Therefore, 
the significant up-conversion is caused by 
intensive modulation fields 
with high modulation frequency.
This would be advantageous 
for realizing microwave/millimeter-wave sources 
toward the 6th-generation mobile communication system.

It is reported \cite{Zaks2012} that 
electron-hole pairs (excitons)
created by weak near-infrared laser 
in semiconductor quantum wells 
is strongly modulated 
by intensive electric fields by a THz free electron laser,
resulting in asymmetric modulation 
with much larger conversion efficiency 
at higher frequency.
The study in the electric counterpart 
consider microscopic nonlinear dynamics 
of exciton under intensive electric fields.
Similarly in this study, 
we consider nonlinear dynamics
of spin waves (magnons) 
created by carrier microwaves 
in the Py layer, 
which are strongly modulated by 
intensive magnetic fields 
of the modulation microwaves.
Fig. \ref{result}(a)
reveals that
the conversion ratio
depends on the power of the carrier and modulation microwaves,
indicating
a nonlinear process 
by the intense magnetic fields 
in the asymmetric conversion.
Numerical simulation in Fig. \ref{calc} 
demonstrates that
the nonlinear magnetization dynamics
due to the demagnetization field 
is essential to the asymmetrical conversion.
Analogous to 
the frequency conversion
in nonlinear optics,
asymmetrical frequency conversions 
using $\mu$ time-varying metamaterials
is traced back to 
the a higher order magnetic susceptibility
through nonlinear magnetization dynamics, 
for example, $\chi^{(3)}$ and nonlinear $\mu$.
The frequency dependence of 
nonlinear $\mu$ 
results in the asymmetric frequency conversion efficiency. 

Last but not least, 
the present metamaterial
is based on the microwave circuit
so that
integration with 
$\varepsilon$ time-modulation systems 
are feasible using EO modulators or
micro electro-mechanical systems (MEMS) \cite{Huang2020-ab}.
This paves a way to 
investigate spatio-temporal four-dimensional control of light and 
the relativistic phenomena in solid state.
Although nonlinear magnetization dynamics 
and microwave response through nonlinear susceptibility 
have attracted attention in spintronics \cite{Sharma2017-gp, Wu2024-co}, 
the physical mechanism is still unclear.
The present study is thus 
an important step in the nonlinear spintronics.
Furthermore,
as the $\mu$ time-varying metamaterial
is regarded as a weakly coupled magnon-photon coupling system 
in the theoretical consideration,
the present study leads to 
nonlinear phenomena in magnon-polaritons \cite{Mita2024}.

\textit{Conclusion} --
We demonstrate magnetic $\mu$ time-varying metamaterials at GHz frequencies 
using ferromagnetic Py.
FMR in the Py thin film 
excited by the carrier microwaves 
is temporally modulated 
by oscillating Oersted magnetic fields 
of another microwave. 
We observe frequency up-and-down conversion 
of the carrier microwaves 
and verify that 
the conversion is caused by the temporal modulation 
of $\mu$ in the Py layer.
Moreover at $f_{\rm m} =$ 0.6 GHz, 
$A_{\rm up}$ is 5.5 times larger than $A_{\rm down}$.
These experimental results are reproduced well 
via numerical calculation, 
indicating that 
the significant up-conversion efficiency 
is traced back to nonlinear magnetization dynamics 
due to demagnetization fields in the $\mu$ time-varying metamaterials.
The present study paves a way to 
development of microwave / millimeter-wave sources 
toward the 6th-generation mobile communication system, 
realization of space-time modulation of both $\varepsilon$ and $\mu$ in metamaterials for experimental demonstration of relativistic and quantum effects, 
and further investigation in the nonlinear spintronics and the magnon-photon coupling systems.

\textit{Acknowledgements} --
We acknowledge 
K. Uchida, M. V. Nguyen, Y. Huang, and Y. Kanamori 
for their valuable discussion in this work.
N. K., S. Ok., and S. T. also thank 
Network Joint Research Center for Materials and Devices (NJRC).
This work is financially supported by JST- CREST (JPMJCR2102).

%\clearpage

%\bibliographystyle{apsrev4-2}
%\bibliographystyle{planenat}
%\bibliography{refprappl}% Produces the bibliography via BibTeX.
%\bibliographystyle{plane}

\clearpage
\pagebreak
\widetext
\begin{center}
\textbf{\large Supplemental Materials for ``Magnetic Permeability Time-varying Metamaterials at Microwave Frequencies''}
\end{center}

\renewcommand\theequation{S\arabic{equation}}
\renewcommand\thefigure{S\arabic{figure}}
\setcounter{equation}{0} 
\setcounter{figure}{0}

\section{Calculated Oersted fields on microstrip lines}
\begin{figure}[h]
\includegraphics[width=8truecm,clip]{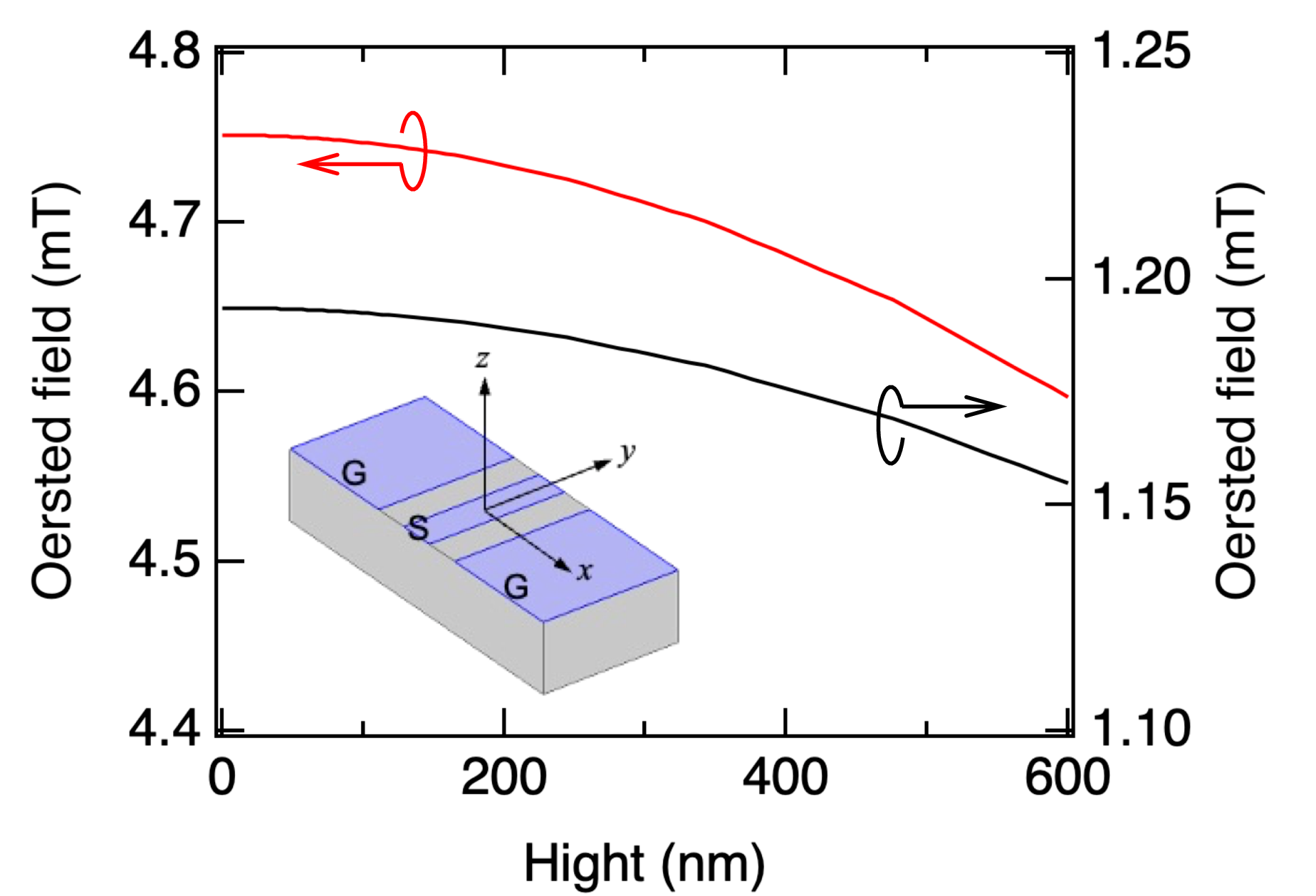}
\caption{
Oersted magnetic fields 
induced by the current flowing 
in the carrier or modulation lines
are plotted as a function of distance from the line surface.
Red (left axis) and black (right axis) lines 
correspond to the calculated Oersted fields 
by modulation microwaves with 20 dBm
and
by carrier microwaves with 8 dBm, 
respectively.
}
\label{cpwfield}
\end{figure}

  The Oersted field
  is evaluated by numerical simulation
  using COMSOL Multiphysics.
  We model coplanar waveguide with
  a 5 {\textmu}m-wide signal (S) line and
  gaps between the S and the ground (G) lines of 5 {\textmu}m.
  Microwaves with 20 dBm ((100 mW) for modulation wave
  and 
  8 dBm (6.31 mW) for carrier wave 
  are applied in the lines.
  Figure S\ref{cpwfield} 
  shows the calculated Oersted magnetic fields 
  along the $y$-direction 
  as a function of the distance 
  from the line surface. 
  Red (left axis) and black (right axis) lines 
  correspond to the calculated Oersted fields 
  by modulation and carrier microwaves, respectively.

\section{Higher-order converted waves}

\begin{figure}[tb!]
\includegraphics[width=15truecm,clip]{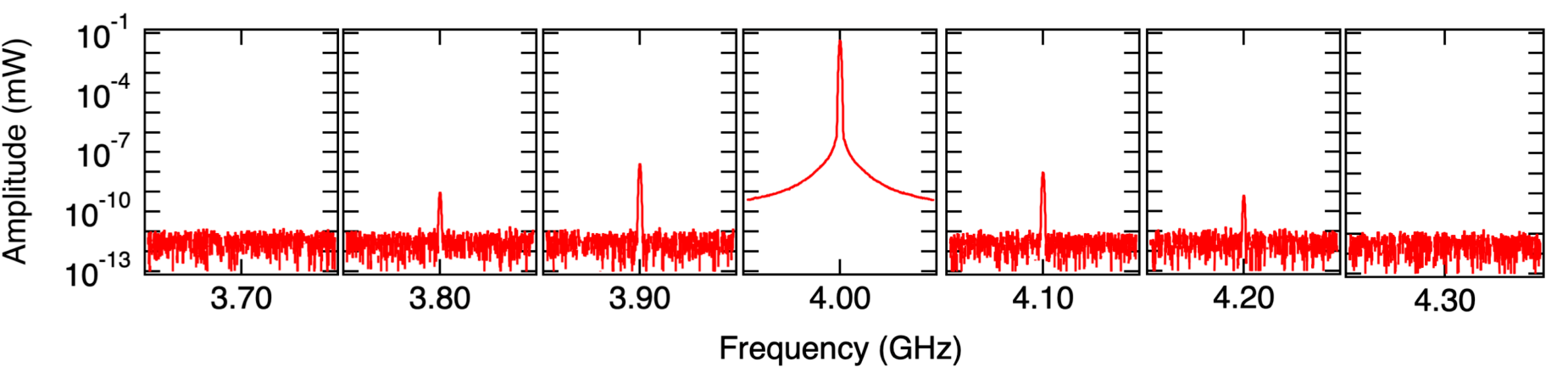}
\caption{
The transmitted microwave spectra of $f_{\rm c} =$ 4.0 GHz
at $f_{\rm m} =$ 0.1 GHz in high sensitivity measurements. 
$\mu_{0}H_{\rm dc} =$ 24.3 mT is applied.
}
\label{conversion2nd}
\end{figure}

  Figure S\ref{conversion2nd} 
  shows 
  the output microwave spectra
  of $f_{\rm c}$= 4.0 GHz
  at $f_{\rm m}$= 0.1 GHz
  under $\mu_0 H_{\rm dc}$ = 24.3 mT.
  To detect small signals,
  the measurement parameters of
  the spectrum analyzer 
  are optimized;
  measurement span = 1 MHz,
  number of data points 2001,
  and resolution band width = 9.1 kHz.
  In addition to 
  the first-ordered conversion signals
  at 4.0 $\pm 0.1$ GHz,
  the second-order signals
  are observed
  at 4.0 $ \pm (2 \times 0.1)$ GHz.
  
\section{CPW-FMR measurements}

\begin{figure}[tb!]
\includegraphics[width=15truecm,clip]{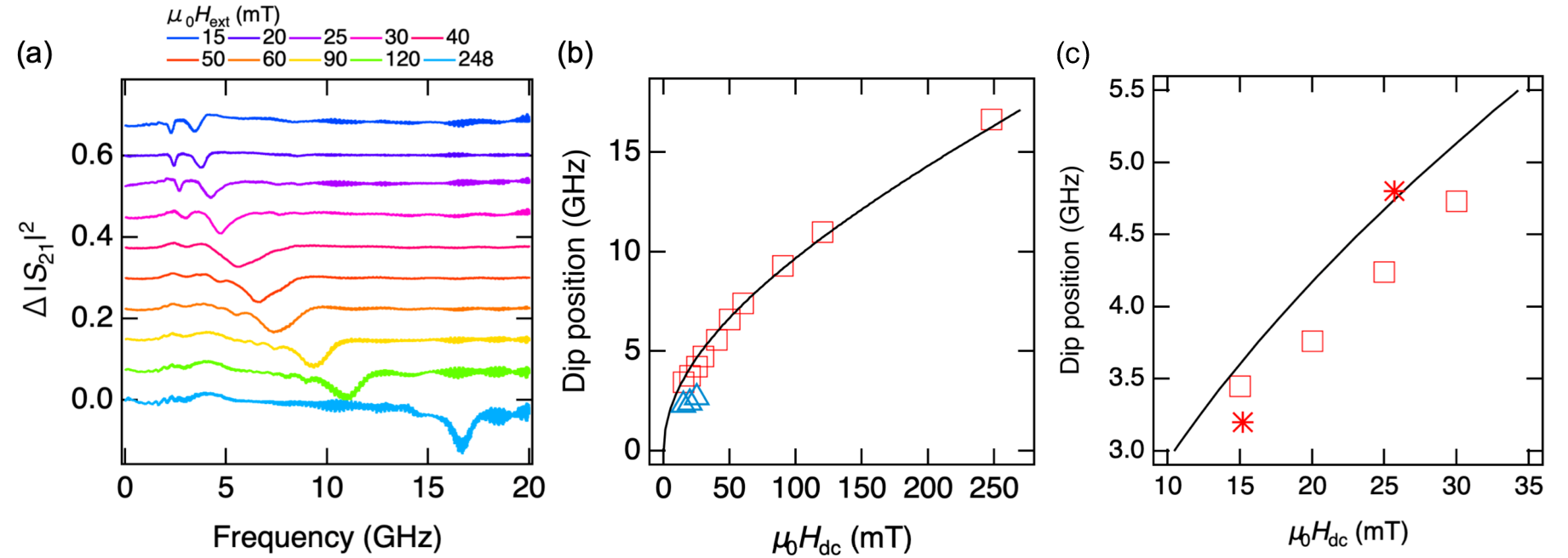}
\caption{
(a) Measured transmission spectra $\Delta |S_{21}|^2$
at various dc magnetic fields $\mu_0 H_{\rm dc}$.
(b) Frequencies of the dip in (a)
are plotted as a function of $\mu_0 H_{\rm dc}$.
Red squares correspond to a higher frequency dip, 
while blue open triangles correspond to a lower one.
The solid line indicates a theoretical curve fitted by the Kittel's formula.
(c) Enlarged view of (b) at a low field.
Red asterisks correspond to the peak field at 3.2 GHz and 4.8 GHz 
in Fig. 3(b) in the main text.}
\label{cpwfmr}
\end{figure}

  For the CPW-FMR measurement, 
  a vector network analyzer 
  is connected to the carrier line, 
  while nothing to the modulation line.
  Frequency spectra of $S$-parameters 
  are obtained 
  at various $\mu_0 H_{\rm dc}$ from 15 to 248 mT.
  Figure S\ref{cpwfmr}(a) shows
  $\Delta |S_{21}|^2 = |S_{21}|^2(\mu_0 H_{\rm dc} \neq 0) - |S_{21}|^2(\mu_0 H_{\rm dc} = 0)$ spectra.
  Here, 
  $|S_{21}|^2(\mu_0 H_{\rm dc} \neq 0)$ 
  is $|S_{21}|^2$ with a specific $\mu_0 H_{\rm dc}$, 
  while $|S_{21}|^2(\mu_0 H_{\rm dc} = 0)$
  with $\mu_0 H_{\rm dc} = 0$.
  The dip in the $\Delta |S_{21}|^2$ 
  correspond to microwave absorption.
  At $\mu_0 H_{\rm dc}$ = 15 mT, 
  dip signals appear at 3.5 GHz and 3.8 GHz.
  As $\mu_0 H_{\rm dc}$ increases, 
  the dip at 3.8 GHz 
  shifts to a higher frequency,
  eventually reaching at 16.7 GHz under $\mu_0 H_{\rm dc} =$ 248 mT.
  On the other hand, 
  the signal at 3.5 GHz under $\mu_0 H_{\rm dc} =$ 15 mT
  becomes week and
  disappears at above $\mu_0 H_{\rm dc} =$ 40 mT. 
  The dip frequencies 
  are plotted as a function of the $\mu_0 H_{\rm dc}$ 
  in Fig. S\ref{cpwfmr}(b).
  The red squares 
  correspond to the dip 
  shifting to a higher frequency 
  as $\mu_0 H_{\rm dc}$ increases.
  The blue triangles 
  correspond to the dip 
  remaining at 3.5 GHz
  and disappearing above 40 mT.
  Fitting the red squares with Eq. (1) in the main text
  yields $\mu_0 M_{\rm eff}$ = 989 mT, 
  which agrees with the experimental results.
  Given that 
  the typical saturation magnetization of Py is 1 T,
  the signals represented by the red squares
  are originated from FMR of Py.
  The resonances dips 
  plotted by blue triangles 
  at a lower frequency
  is likely to be caused by the voltex mode resonance of the Py patch.

\section{Hysteresis of converted signal amplitudes}

\begin{figure}[tb!]
\includegraphics[width=15truecm,clip]{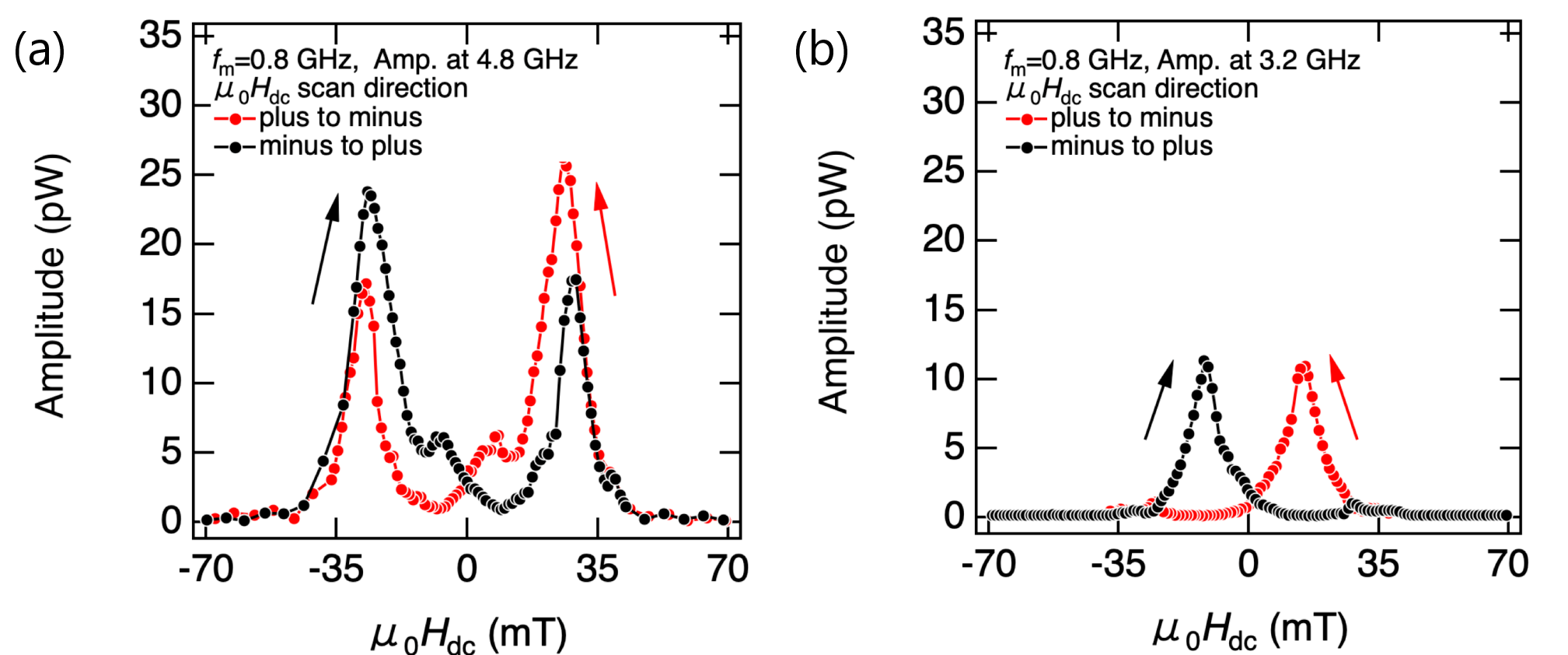}
\caption{(a) $A_{\rm up}$ at 4.8 GHz 
and (b) $A_{\rm down}$ 3.2 GHz 
with $f_{\rm c}$ = 4.0 GHz and $f_{\rm m}$ = 0.8 GHz 
are plotted as a function of $\mu_{0} H_{\rm dc}$.
Red and black color correspond to 
the field-sweep direction 
from +70 to -70 mT 
and from -70 to + 70 mT, 
respectively.
}
\label{hysteresis}
\end{figure}

  Hysteresis 
  of the amplitude of the up- and down-converted signal 
  with $f_{\rm m}$ = 0.8 GHz
  is investigated.
  In Figs. S\ref{hysteresis}(a) and S\ref{hysteresis}(b), 
  $A_{\rm up}$ at 4.8 GHz 
  and $A_{\rm down}$ at 3.2 GHz 
  with $f_{\rm c}$ = 4.0 GHz and $f_{\rm m}$ = 0.8 GHz 
  are plotted as a function of $\mu_{0} H_{\rm dc}$, respectively.
  Red and black colors correspond to the field-sweep direction 
  from +70 to $-$70 mT and from $-$70 to + 70 mT, 
  respectively.  
  When $\mu_{0} H_{\rm dc}$ changes from +70 to $-$70 mT, 
  red circles in Fig. S\ref{hysteresis}(a) 
  corresponding to $A_{\rm up}$
  shows two peaks 
  at $\mu_{0} H_{\rm dc} = $ 25.7 mT and $-$ 27.2 mT
  as mentioned in the main text.
  The amplitude at 25.7 mT
  is larger than that at $-$ 27.2 mT.
  On the other hand,
  when the sweep direction is reversed from $-$70 to +70 mT
  (black circles), 
  $A_{\rm up}$ shows peaks at  $-$26.7 mT and 29.1 mT.
  The variation of $A_{\rm up}$, thus,
  shows hysteresis depending on sweep direction of $\mu_{0} H_{\rm dc}$.
  $A_{\rm down}$ at 3.2 GHz in Fig. S\ref{hysteresis}(b) 
  present a similar hysteresis feature. 
  This indicates that 
  the magnetic domain structure 
  is a non-uniform pattern, 
  for example, a vortex state. 

\section{Details of theoretical model}
To describe the experiment, 
a forced oscillation model 
based on Maxwell's equation and Landau-Lifshitz-Gilbert (LLG) equation 
is used, 
where the microwave magnetic field (photon) is weekly coupled 
with the magnetization dynamics (magnon). 
The microwave magnetic field 
${\bf H}_{\rm ac}^{\rm c}(t) = H_{\rm ac}^{\rm c}(t)\hat{\bf y}$ 
with an angler frequency $\omega$ 
is governed by the following equation of motion 
based on Maxwell's equations \cite{MitaPRAP2025}:
\begin{align}
\left(  \frac{d^2}{dt^2} + 2\beta\omega\frac{d}{dt} + \omega^2\right){\bf H}_{\rm ac}^{\rm c}
= -\eta_V\mu_0M_{\rm s}\frac{d^2}{dt^2}m_y\hat{\bf y}+\omega_{\rm r}^2{\bf H}_{\rm rf}(t),
\label{Maxwell}
\end{align}
where $\beta$ is an effective damping of the microwave magnetic field, 
$\mu_{0}$ is the permeability of vacuum, 
$M_{s}$ is the saturation magnetization of a ferromagnet, 
and ${\bf H}_{\rm rf}(t) = H_{\rm rf}\cos{(\omega_{\rm r}t)}\hat{\bf y}$ 
is a driving magnetic field 
generated by carrier waves
with an angler frequency $\omega_{\rm r}$. 
Here, $\eta_V$ represents the effective volume ratio 
between the microwave magnetic field and the ferromagnet.
On the other hand, 
the magnetization dynamics for a uniform magnon mode 
are described by the LLG equation:
\begin{align}
\frac{d \bf m}{d t} = -\gamma{\bf m}\times\left[  -\frac{1}{M_{\rm s}}\frac{\delta U_{\rm m}}{\delta{\bf m}} + \left(  {\bf H}_{\rm ac}^{\rm c} + {\bf H}_{\rm ac}^{\rm m}\right)\right] + \alpha{\bf m}\times\frac{d \bf m}{d  t},
\label{LLG}
\end{align}
where ${\bf m}$ is the unit vector 
aligned with the magnetization direction of the ferromagnet, 
$\gamma$ is the gyromagnetic ratio, 
and $\alpha$ is the intrinsic Gillbert damping constant. 
Also, 
${\bf H}_{\rm ac}^{\rm m}(t) = H_{\rm ac}^{\rm m}\cos{(\omega_{\rm m}t)}\hat{\bf x}$ 
in Eq.~$\eqref{LLG}$ represents the modulation Oersted field.

Here, the total magnetic energy is as follows
\begin{align}
U_{\rm m} = - \mu_0M_{\rm s}{\bf m}\cdot{\bf H}_{\rm dc} + \frac{1}{2}\mu_0M_{\rm s}^2m_z^2,
\label{magneticenergy}
\end{align}
where ${\bf H}_{\rm dc}=H_{\rm dc}\hat{\bf x}$ is a dc external magnetic field. 
Note that the second term in Eq.~$\eqref{magneticenergy}$ 
represents the demagnetizing field of the magnetic thin film. 
We numerically solve the coupled Eqs.~$\eqref{Maxwell}$  and $\eqref{LLG}$ 
to evaluate the modulated microwave spectrum, 
$|\tilde{h}_{\rm ac}^{\rm c}|$, 
which corresponds to the Fourier spectrum of ${\bf H}_{\rm ac}^{\rm c}/M_{\rm s}$. 
For calculations in the main text, 
we use parameters: 
$\mu_{0}H_{\rm dc} = 25$~mT, 
$\mu_{0}H_{\rm rf} = 1.15$~mT, 
$\mu_{0}H_{\rm ac}^{\rm m} = 5$~mT, 
$\omega_{\rm r}/(2\pi) = \omega/(2\pi) = 4~{\rm GHz}$, 
$\eta_V = 0.01$, 
$\beta = 0.1$, 
$\alpha = 0.01$, 
$\mu_0M_{\rm s} = 0.8$~T, 
and $\gamma = 1.76\times10^{11}$~/Ts.

%\bibliographystyle{apsrev4-2}
%\bibliographystyle{planenat}
%\bibliography{refprappl}% Produces the bibliography via BibTeX.

\begin{thebibliography}{}\label{sec:TeXbooks}

\bibitem{Noether} Y. Kosmann-Schwarzbach, B. E. Schwarzbach , {\it The Noether Theorems: Invariance and Conservation Laws in the Twentieth Century} (Springer, New York, 2011).
\bibitem{Hecht} E. Hecht, {\it Optics} (Pearson Education Limited, Harlow, 2013).
\bibitem{Miyamaru2021-ph} F. Miyamaru et al., Ultrafast Frequency-Shift Dynamics at Temporal Boundary Induced by Structural-Dispersion Switching of Waveguides, Phys. Rev. Lett. {\bf 127}, 053902 (2021).
\bibitem{Udem2002} T. Udem, R. Holzwarth, T. H\"{a}nsch, Nature {\bf 416}, 233 (2002).
\bibitem{Ishizawa2023} A. Ishizawa, T. Nishikawa, K. Hitachi, T. Akatsuka, K. Oguri, Sci. Rep. {\bf 13}, 8750 (2023).
\bibitem{Shelby2001-rn} R. A. Shelby, D. R. Smith, and S. Schultz, Experimental verification of a negative index of refraction, Science {\bf 292}, 77 (2001).
\bibitem{Schurig2006-kk} D. Schurig, J. J. Mock, B. J. Justice, S. A. Cummer, J. B. Pendry, A. F. Starr, and D. R. Smith, Metamaterial electromagnetic cloak at microwave frequencies, Science {\bf 314}, 977 (2006).
\bibitem{Galiffi2022-kc} E. Galiffi, R. Tirole, S. Yin, H. Li, S. Vezzoli, P. A. Huidobro, M. G. Silveirinha, R. Sapienza, A. Al\'{u}, and J. B. Pendry, Photonics of time-varying media, Advanced Photonics {\bf 4}, 0140022 (2022).
\bibitem{Zhang2022-ei}
L. P. Zhang, H. C. Zhang, J. Zhang, L. Y. Niu, P. H. He, C. Wei, W. Tang, and T. J. Cui, Reprogrammable control of electromagnetic spectra based on time-coding plasmonic metamaterials, Appl. Phys. Lett. {\bf 121}, 161702 (2022).
\bibitem{Taravati2021-wm}
S. Taravati and G. V. Eleftheriades, Pure and Linear Frequency-Conversion Temporal Metasurface, Phys. Rev. Appl. {\bf 15}, 064011 (2021).
\bibitem{Guo2019-vq}
X. Guo, Y. Ding, Y. Duan, and X. Ni, Nonreciprocal metasurface with space-time phase modulation, Light Sci. Appl. {\bf 8}, (2019).
\bibitem{Pacheco-Pena2020-yq} 
V. Pacheco-Pe\~{n}a and N. Engheta, Antireflection temporal coatings, Optica {\bf 7}, 323 (2020).
\bibitem{Liberal2023-ld} 
I. Liberal, J. E. V\'{a}zquez-Lozano, and V. Pacheco-Pe\~{n}a, Quantum antireflection temporal coatings: Quantum state frequency shifting and inhibited thermal noise amplification, Laser Photon. Rev. {\bf 17}, (2023).
\bibitem{Zhou2020-of} Y. Zhou, M. Z. Alam, M. Karimi, J. Upham, O. Reshef, C. Liu, A. E. Willner, and R. W. Boyd, Broadband frequency translation through time refraction in an epsilon-near-zero material, Nat. Commun. {\bf 11}, 1 (2020).
\bibitem{Apffel2022-go}
B. Apffel and E. Fort, Frequency Conversion Cascade by Crossing Multiple Space and Time Interfaces, Phys. Rev. Lett. {\bf 128}, 064501 (2022).
\bibitem{Moussa2023-pv} 
H. Moussa, G. Xu, S. Yin, E. Galiffi, Y. Ra'di, and A. Al\'{u}, Observation of temporal reflection and broadband frequency translation at photonic time interfaces, Nat. Phys. {\bf 1} (2023).
\bibitem{Liu2023-nw}
T. Liu, J.-Y. Ou, K. F. MacDonald, and N. I. Zheludev, Photonic metamaterial analogue of a continuous time crystal, Nat. Phys. {\bf 19}, 986 (2023).
\bibitem{Pacheco-Pena2020-jn} 
V. Pacheco-Pe\~{n}a and N. Engheta, Temporal aiming, Light Sci. Appl. {\bf 9}, 129 (2020).
\bibitem{Huidobro2019-eg} P. A. Huidobro, E. Galiffi, S. Guenneau, R. V. Craster, and J. B. Pendry, Fresnel drag in space-time-modulated metamaterials, Proc. Natl. Acad. Sci. USA {\bf 116}, 24943 (2019).
\bibitem{Engheta2023-ta}
N. Engheta, Four-dimensional optics using time-varying metamaterials, Science {\bf 379}, 1190 (2023).
\bibitem{Kort-Kamp2021-pa} 
W. J. M. Kort-Kamp, A. K. Azad, and D. A. R. Dalvit, Space-time quantum metasurfaces, Phys. Rev. Lett. {\bf 127}, 043603 (2021).
\bibitem{Pendry2024-pb} 
J. Pendry and S. A. Horsley, QED in space-time varying materials, APL Quantum {\bf 1}, (2024).
\bibitem{Silveirinha2023-ip} 
M. G. Silveirinha, Hawking-type radiation in transluminal gratings, Proc. Natl. Acad. Sci. U. S. A. {\bf 120}, e2313369120 (2023).
\bibitem{Liu2020-rk} 
B. Liu, H. Giddens, Y. Li, Y. He, S.-W. Wong, and Y. Hao, Design and experimental demonstration of Doppler cloak from spatiotemporally modulated metamaterials based on rotational Doppler effect, Opt. Express {\bf 28}, 3745 (2020).
\bibitem{Kodama2023PRAppl} 
T. Kodama, N. Kikuchi,S. Okamoto, S. Ohno, and S. Tomita, Spin-current Driven Permeability Variation for Time-varying Magnetic Metamaterials, Phys. Rev. Appl. {\bf 19}, 044080 (2023).
\bibitem{Kodama2024PRB}  
T. Kodama, N. Kikuchi, T. Chiba, S. Okamoto, S. Ohno, and S. Tomita, Direct observation of current-induced nonlinear spin torque in Pt-Py bilayers, Phys. Rev. B. {\bf 109}, 214419 (2024).
\bibitem{Oka2019} T. Oka and S. Kitamura, Floquet Engineering of Quantum Materials, Annu. Rev. Condens. Matter Phys. {\bf 10}, 387 (2019).
\bibitem{Yin2022} S. Yin, E. Galiffi, and A. Al\'{u}, Floquet metamaterials, eLight {\bf 2}, 8 (2022). 
\bibitem{Sharma2017-gp} 
R. Sharma, N. Sisodia, E. Iacocca, A. A. Awad, J. \r{A}kerman, and P. K. Muduli, A high-speed single sideband generator using a magnetic tunnel junction spin torque nano-oscillator, Sci. Rep. {\bf 7}, 13422 (2017).
\bibitem{Wu2024-co} 
J. Wu, J. Liu, Z. Ren, M. Y. Leung, W. K. Leung, K. O. Ho, X. Wang, Q. Shao, and S. Yang, Wideband coherent microwave conversion via magnon nonlinearity in a hybrid quantum system, Npj Spintronics {\bf 2}, 1 (2024).
\bibitem{Zheng2023} S. Zheng, Z. Wang, Y. Wang, F. Sun, Q. He, P. Yan, H. Y. Yuan, Tutorial: Nonlinear magnonics, J. Appl. Phys. {\bf 134}, 151101 (2023).
\bibitem{SM} 
See Supplemental Materials at URL for Oersted fields on microstrip lines, experimentally obtained higher-order converted waves, co-planar waveguide FMR measurement, hysteresis feature of converted waves, and Details of theoretical model. 
\bibitem{Wurft2019-zu}  
T. Wurft, W. Raberg, K. Pr\"{u}gl, A. Satz, G. Reiss, and H. Br\"{u}ckl, Evolution of magnetic vortex formation in micron-sized disks, Appl. Phys. Lett. {\bf 115}, 132407 (2019).
\bibitem{Chiba2024-in}  
T. Chiba, T. Komine, and T. Aono, Ultrastrong-coupled magnon-polariton in a dynamical inductor based on magnetic-insulator/topological-insulator bilayers, Appl. Phys. Lett. {\bf 124}, 012402 (2024).
\bibitem{Mita2024}  
K. Mita, T. Chiba, T. Kodama, T. Ueda, T. Nakanishi, K. Sawada, and S. Tomita, Ultrastrongly coupled and directionally nonreciprocal magnon polaritons in magnetochiral metamolecules, Phys. Rev. Applied {\bf 23}, L011004 (2025).
\bibitem{Zaks2012} B. Zaks, R. B. Liu, and M. S. Sherwin, Experimental observation of electron-hole recollisions, Nature  {\bf 483}, 580 (2012)
\bibitem{Huang2020-ab} Y. Huang, K. Nakamura, Y. Takida, H. Minamide, K. Hane, and Y. Kanamori, Actively tunable THz filter based on an electromagnetically induced transparency analog hybridized with a MEMS metamaterial, Sci. Rep. {\bf 10}, 20807 (2020).


\end{thebibliography}

\begin{thebibliography}{99}\label{sec:TeXbooks}
\bibitem[6]{MitaPRAP2025} 
K. Mita, T. Chiba, T. Kodama, T. Ueda, T. Nakanishi, K. Sawada, and S. Tomita, Ultrastrongly coupled and directionally nonreciprocal magnon polaritons in magnetochiral metamolecules, Phys. Rev. Applied {\bf 23}, L011004 (2025).
\end{thebibliography}
%\bibliographystyle{plane}

\end{document}